\documentclass[conference]{IEEEtran}
\IEEEoverridecommandlockouts
\usepackage{cite}
\usepackage{amsmath,amssymb,amsfonts}
\usepackage{algorithmic}
\usepackage{graphicx}
\usepackage{textcomp}
\usepackage{xcolor}
\usepackage{placeins}
\def\BibTeX{{\rm B\kern-.05em{\sc i\kern-.025em b}\kern-.08em
    T\kern-.1667em\lower.7ex\hbox{E}\kern-.125emX}}
\begin{document}

\title{Automatic Cranial Defect Reconstruction with Self-Supervised Deep Deformable Masked Autoencoders \\
\thanks{The project was funded by The National Centre for Research and Development, Poland under Lider Grant no: LIDER13/0038/2022 (DeepImplant). We gratefully acknowledge Polish HPC infrastructure PLGrid support within computational grant no. PLG/2023/016239. The research was partially supported by the program "Excellence Initiative - Research University" for AGH University.}
\thanks{$^{1}$Marek Wodzinski, Daria Hemmerling, and Mateusz Daniol are with the AGH University of Krakow, Department of Measurement and Electronics, Krakow, Poland.}
\thanks{$^{2}$Marek Wodzinski is also with the University of Applied Sciences Western Switzerland (HES-SO Valais), Institute of Informatics, Sierre, Switzerland.

© 2024 IEEE must be obtained for all other uses, in any current or future media, including reprinting/republishing this material for advertising or promotional purposes, creating new collective works, for resale or redistribution to servers or lists, or reuse of any copyrighted component of this work in other works.}
}

\author{Marek Wodzinski$^{1,2}$, Daria Hemmerling$^{1}$, Mateusz Daniol$^{1}$
}

\maketitle

\begin{abstract}

Thousands of people suffer from cranial injuries every year. They require personalized implants that need to be designed and manufactured before the reconstruction surgery. The manual design is expensive and time-consuming leading to searching for algorithms whose goal is to automatize the process. The problem can be formulated as volumetric shape completion and solved by deep neural networks dedicated to supervised image segmentation. However, such an approach requires annotating the ground-truth defects which is costly and time-consuming. Usually, the process is replaced with synthetic defect generation. However, even the synthetic ground-truth generation is time-consuming and limits the data heterogeneity, thus the deep models' generalizability. In our work, we propose an alternative and simple approach to use a self-supervised masked autoencoder to solve the problem. This approach by design increases the heterogeneity of the training set and can be seen as a form of data augmentation. We compare the proposed method with several state-of-the-art deep neural networks and show both the quantitative and qualitative improvement on the SkullBreak and SkullFix datasets. The proposed method can be used to efficiently reconstruct the cranial defects in real time.

\end{abstract}

\begin{IEEEkeywords}
Deep Learning, Cranial Implant, Shape Completion, Skull Defect, Masked Autoencoder, Self-Supervised Learning
\end{IEEEkeywords}

\section{Introduction}

Cranial injuries are a common result of neurosurgery, traffic accidents, or warfare. Thousands of people suffer from such injuries every year and require personalized implants~\cite{global}. The process of modeling and manufacturing such implants requires expertise, is time-consuming, and leads to substantial costs and waiting time. However, the process can be at least partially automated by deep learning algorithms~\cite{autoimplant1,autoimplant2}.

\begin{figure}[!htb]
    \centering
    \includegraphics[width = 0.5\textwidth]{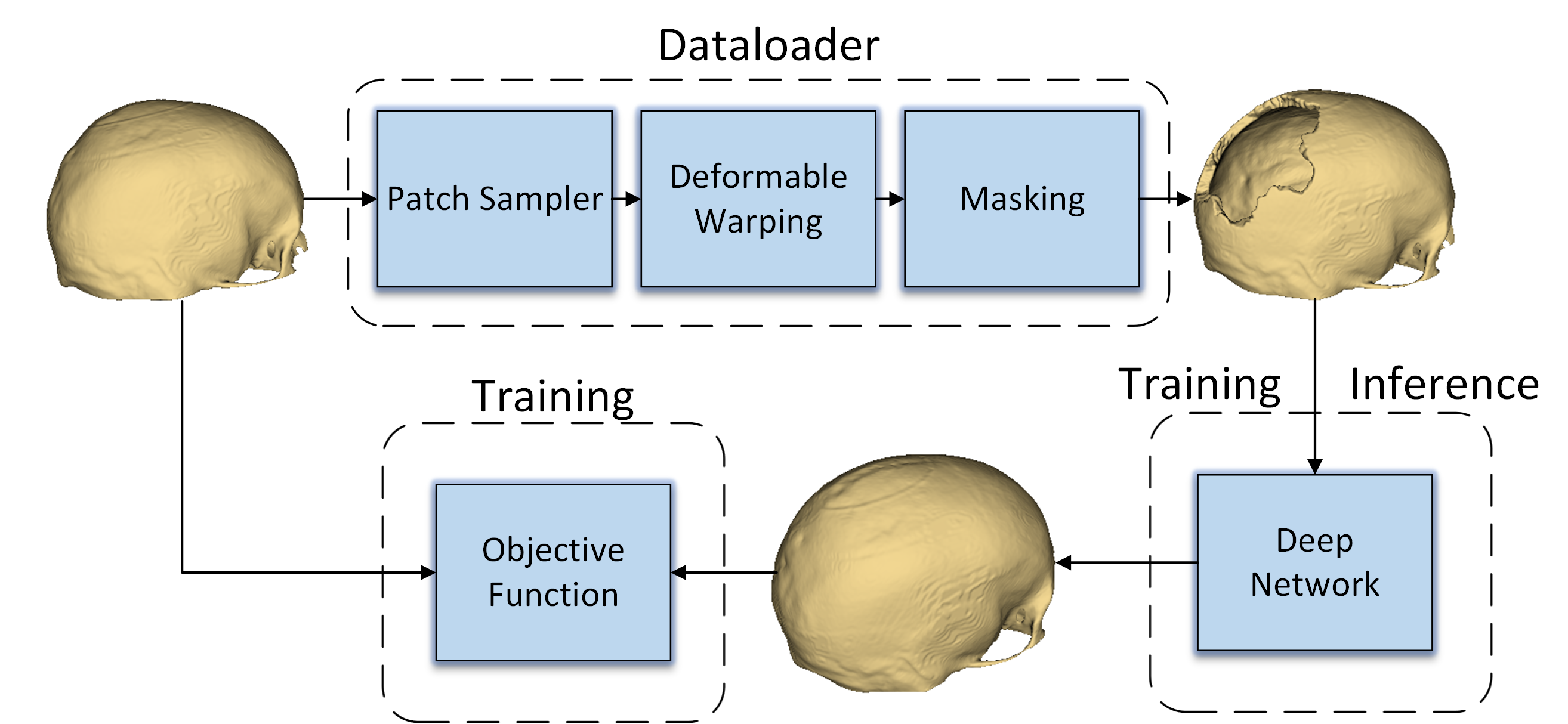}
    \caption{The pipeline of the proposed deep deformable masked autoencoder. Firstly, The healthy skull is loaded, then masking patches are randomly selected, deformably transformed, and used for masking to produce the defective skull. Then, a dedicated encoder-decoder network is used to perform the reconstruction.}
    \label{fig:pipeline}
\end{figure}

Nowadays, the most common approach to automatic cranial defect reconstruction is to treat it as a volumetric segmentation that can be solved by deep neural networks~\cite{autoimplant1,autoimplant2}. Numerous works present the usefulness of segmentation networks in this context, varying from the use of simple convolutional architectures~\cite{yang,mahdi,wodzinski1}, to more advanced methods involving image registration-based or generative augmentation~\cite{ellis1,kwarciak1,wodzinski2}. Some of the works attempt to reformulate the problem and to solve it in other domains, e.g. as a point-cloud completion task~\cite{wodzinski3,friedrich1}. Nevertheless, the major problem that all the approaches attempt to solve is to increase the heterogeneity of the training set leading to the improvement of the model generalizability. The challenge comes from the fact that it is extremely costly and time-consuming to acquire and annotate cases with real cranial defects. Therefore, the broadly used open datasets like SkullBreak, SkullFix, or MUG500 consist mostly of real skulls with synthetic defects suffering from relatively large homogeneity and difficulties with generalization into real cases~\cite{skullbreak,mug500}. Therefore, it would be beneficial to propose a method that automatically generates heterogeneous ground-truth defects. The masked autoencoders (MAEs) seem to be perfect for this task.

The idea behind the masker autoencoders is to randomly delete part of the input and then train an encoder-decoder network to reconstruct the missing data. The task is considerably more difficult than training the classical autoencoders to just recover its input. The self-supervisedly pretrained masked autoencoders are then useful for other downstream tasks because they learn both general and detailed features associated with the data. Since masked autoencoders do not require the ground-truth annotations, they can be seen as a powerful self-supervised pretraining tool~\cite{he2022masked,feichtenhofer2022masked,zhang2022survey}.

\textbf{Contribution: } In this work, we propose an alternative approach to the automatic cranial defect reconstruction by using a deep deformable masked autoencoder. We prove its usability and compare it to the state-of-the-art deep architectures dedicated to volumetric segmentation. We confirm that the implicit heterogeneity introduced by the masked autoencoder training improves the automatic cranial defect reconstruction.

\section{Methods}

\subsection{Deformable Masked Autoencoder}

In this work, we propose an approach based on a deep masked autoencoder. During training, for each case, we generate a random number of patches with variable sizes that are going to mask out part of the healthy skulls, thus generating defective ones. The generated patches are randomly transformed by deformable elastic deformation. It results in smooth shapes that resemble real cranial defects. Such patches are then used to generate the ground truth for the self-supervision.

\begin{figure}[!htb]
    \centering
    \includegraphics[width = 0.5\textwidth]{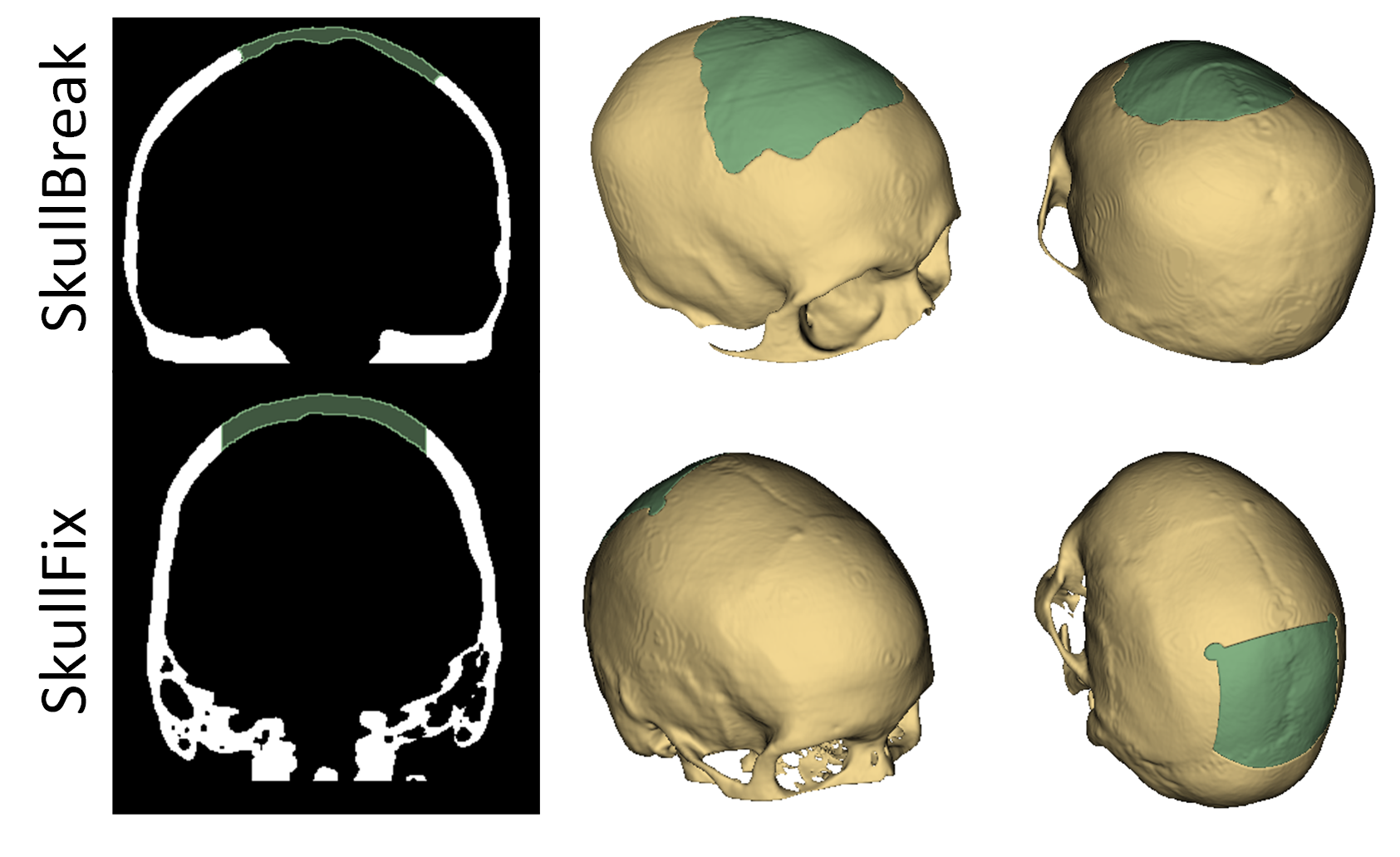}
    \caption{Exemplary cases from the SkullFix and SkullBreak datasets. Note the sharp boundaries in the SkullFix dataset, and the smooth, irregular ones in the SkullBreak dataset.}
    \label{fig:example_data}
\end{figure}

The defective inputs are then passed to the masked autoencoder network. The goal of the network is to reconstruct the input shape from the defective ones. This way, we slightly reformulate the problem from the missing shape segmentation into learning the overall shape of skulls. Since the patches are generated, inserted, and deformed randomly, this approach strongly increases the data heterogeneity, thus leading to improvements in the model's generalizability. In this work, we use a Residual 3-D UNet as the autoencoder backbone~\cite{wodzinski1}. The processing pipeline is shown in Figure~\ref{fig:pipeline}.

\begin{figure*}[!htb]
    \centering
    \includegraphics[width = 0.7\textwidth]{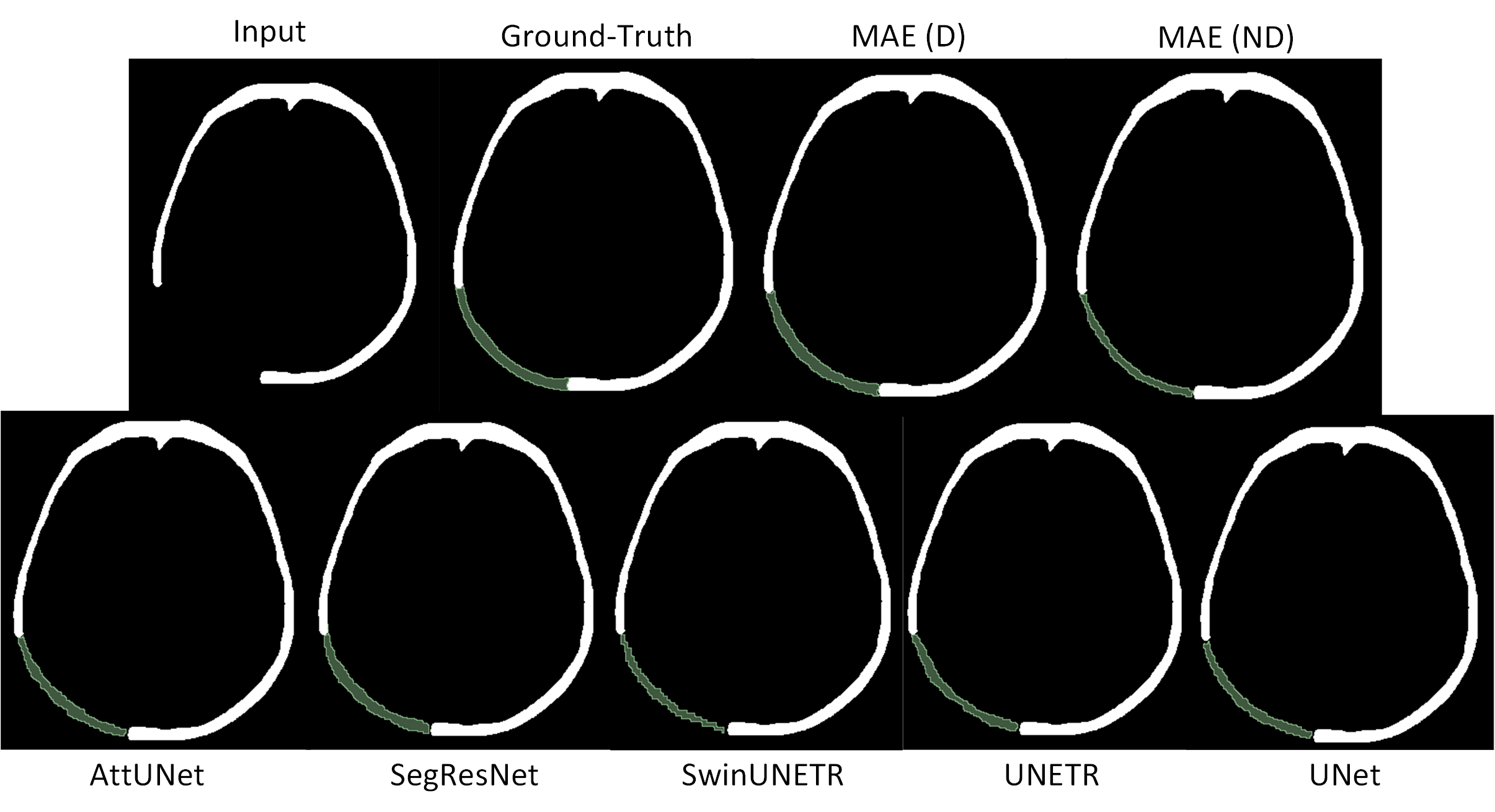}
    \caption{Exemplary visual comparison of the proposed deformable MAE to the volumetric segmentation methods.}
    \label{fig:vis_comp}
\end{figure*}

Since the processed data is binary, we reconstruct the patches using the Soft Dice Score as the loss function. The initial ablations confirmed that such an approach is more stable and converges faster than experiments using mean absolute or mean squared differences. Other widely used objective functions to train masked autoencoders in the computer vision domain are not useful in the discussed context since we are processing directly binary data.

\subsection{Datasets and Experimental Setup}

\begin{figure*}
    \centering
    \includegraphics[width = 0.92\textwidth]{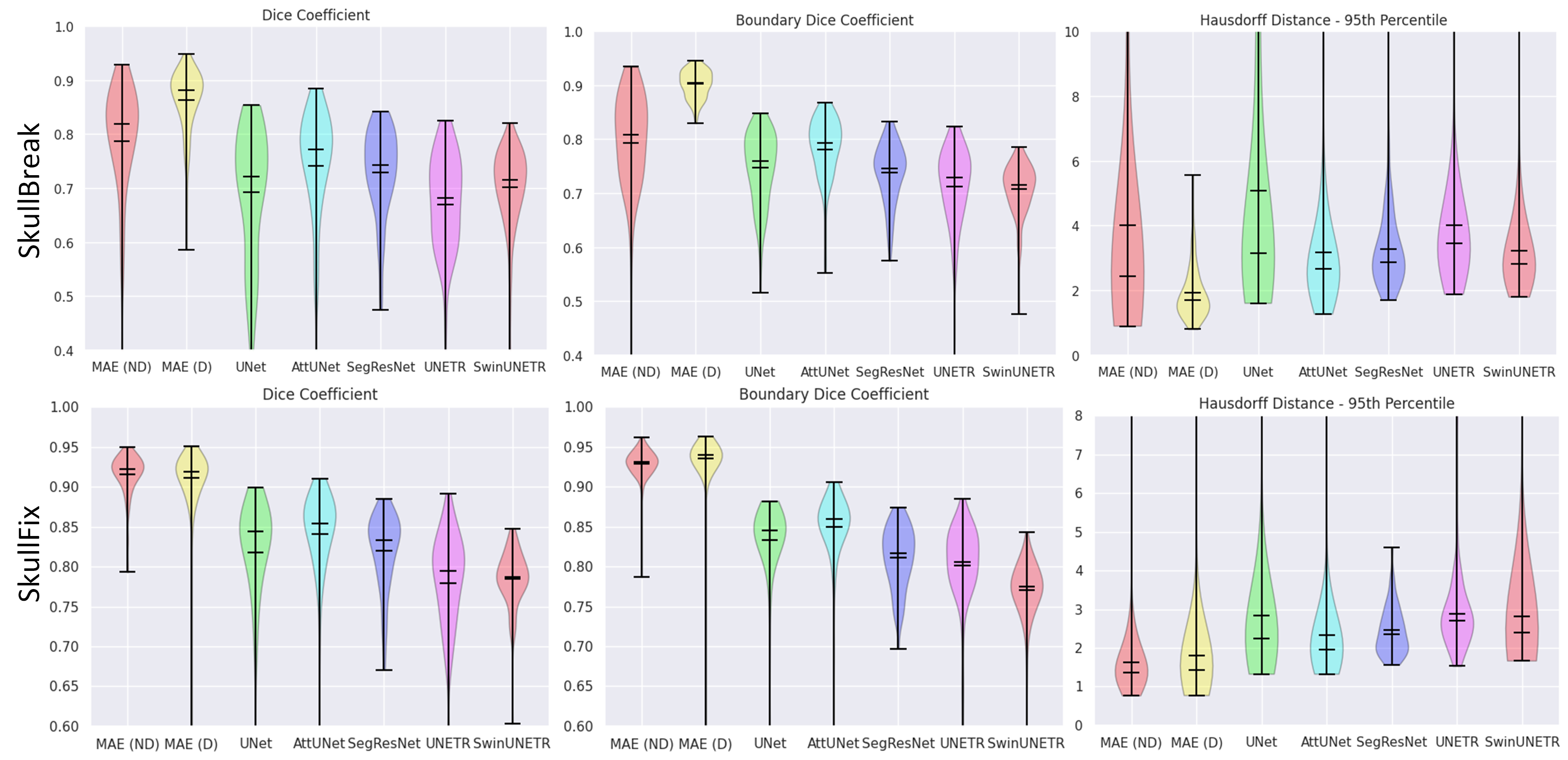}
    \caption{The qualitative results of the proposed approach compared to the state-of-the-art segmentation networks. The "D" denotes the proposed approach with the deformable elastic deformations and "ND" is the masking with sharp edges. Note the significant improvement of the proposed approach for the SkullBreak dataset.}
    \label{fig:results}
\end{figure*}

\begin{table*}[!htb]
\centering
\caption{Table presenting the mean and median value for DSC, BDSC, and HD95 for the proposed MAE-based approach, and the state-of-the-art networks in volumetric segmentation. The "D" denotes the proposed approach with the deformable elastic deformations and "ND" masking with sharp edges.}
\renewcommand{\arraystretch}{1.0}
\footnotesize
\resizebox{0.90\textwidth}{!}{%
\begin{tabular}{lcccccccccccc}
\label{tab:results}
Method & \multicolumn{6}{c}{SkullBreak} & \multicolumn{6}{c}{SkullFix} \tabularnewline
\hline
\multicolumn{1}{c}{} & \multicolumn{2}{c}{DSC $\uparrow$} & \multicolumn{2}{c}{BDSC $\uparrow$} & \multicolumn{2}{c}{HD95 [mm] $\downarrow$} & \multicolumn{2}{c}{DSC $\uparrow$} & \multicolumn{2}{c}{BDSC $\uparrow$} & \multicolumn{2}{c}{HD95 [mm] $\downarrow$} \tabularnewline

\multicolumn{1}{c}{} & Avg & Med & Avg & Med & Avg & Med & Avg & Med & Avg & Med & Avg & Med \tabularnewline

MAE (ND) & 0.787 & 0.820 & 0.793 & 0.808 & 4.02 & 2.43 & \textbf{0.916} & \textbf{0.922} & 0.929 & 0.930 & \textbf{1.63} & \textbf{1.36}
\tabularnewline

MAE (D) & \textbf{0.863} & \textbf{0.882} & \textbf{0.902} & \textbf{0.905} & \textbf{1.92} & \textbf{1.70} & 0.911 & 0.919 & \textbf{0.935} & \textbf{0.939} & 1.80 & 1.43
\tabularnewline

UNet & 0.692 & 0.721 & 0.746 & 0.758 & 5.10 & 3.15 & 0.817 & 0.844 & 0.832 & 0.845 & 2.83 & 2.24
\tabularnewline

AttUnet & 0.741 & 0.772 & 0.781 & 0.793 & 3.17 & 2.65 & 0.841 & 0.854 & 0.850 & 0.859 & 2.33 & 1.94
\tabularnewline

SegResNet & 0.729 & 0.743 & 0.737 & 0.746 & 3.28 & 2.87 & 0.819 & 0.833 & 0.811 & 0.816 & 2.45 & 2.34
\tabularnewline

UNETR & 0.670 & 0.681 & 0.712 & 0.728 & 4.01 & 3.44 & 0.779 & 0.794 & 0.801 & 0.806 & 2.88 & 2.71
\tabularnewline

SwinUNETR & 0.702 & 0.715 & 0.708 & 0.715 & 3.21 & 2.82 & 0.785 & 0.786 & 0.770 & 0.775 & 2.82 & 2.39
\tabularnewline

\hline
\end{tabular}}

\end{table*}

We train the proposed method on the healthy skulls available in the SkullFix and SkullBreak datasets~\cite{skullbreak}. The training part of the SkullFix and SkullBreak datasets consists of 100 and 114 skulls respectively, resulting in 214 training cases. The volumetric segmentation networks used for comparison are trained using the combined datasets with known ground truth, resulting in a training set consisting of 570 defective cases. Exemplary cases from the SkullFix and SkullBreak datasets are shown in Figure~\ref{fig:example_data}.

The proposed method, as well as the methods used for comparison, are evaluated using the SkullBreak and Skull Fix test sets consisting of 100 and 110 skulls with defects respectively. We compare the proposed method to several state-of-the-art segmentation architectures: (i) the Resiudal UNet~\cite{ronneberger2015u,kerfoot2019left}, (ii) the UNETR~\cite{hatamizadeh2022unetr}, (iii) the SwinUNETR~\cite{hatamizadeh2021swin}, (iv) the Attention UNet~\cite{oktay2018attention}, (v) the SegResNet~\cite{myronenko20193d}, all openly available in the MONAI library~\cite{Cardoso_MONAI_An_open-source_2022}. We also perform ablation studies related to the influence of random deformable transformations applied during the input masking.

All the experiments are implemented in PyTorch with the support of TorchIO library~\cite{torchio}. The input cases are centered, cropped to the skull, and resampled to 256x256x256 voxels. The evaluation metrics were calculated on the skull defects using original resolution after upsampling the inference output, without further postprocessing. The skull defects were calculated by morphological operations. We evaluate the methods using the Dice Coefficient (DSC), Boundary Dice Coefficient (BDSC), and 95th percentile of Hausdorff distance (HD95), following the conventions introduced by the authors of the SkullFix and SkullBreak datasets during the AutoImplant challenge~\cite{autoimplant1,autoimplant2}. All the networks were trained until convergence using a computing cluster with NVIDIA A100 40GB GPUs. The Soft Dice Loss was used as the objective function, the AdamW as the optimizer with learning rate/weight decay equal to 0.001 and 0.01 respectively, and with decreasing the learning rate by exponentially decaying scheduler with the ratio equal to 0.995.

\section{Results and Discussion}

The visual comparison of the methods on a randomly chosen case from the test set is shown in Figure~\ref{fig:vis_comp}. The comparison presenting the performance concerning DSC, BDSC, and HD95 are presented in Figure~\ref{fig:results}. The results are summarized in Table~\ref{tab:results}. The figure compare the proposed method to the state-of-the-art medical segmentation architectures. 

The results confirm that the proposed method outperforms the state-of-the-art solutions by a considerable margin. The trained model improves the reconstruction quality with respect to all the quantitative metrics. Moreover, the results show that the random elastic deformations of the generated patches are crucial to improve the performance on the SkullBreak dataset. In case of sharp patches, the network is unable to learn smooth boundaries that are crucial for an accurate defect reconstruction. On the other hand, the effect is less significant on the SkullFix dataset where all the defects are sharp and the deformable improvement is not crucial.

Another advantage of the method is connected with easy extendability into new datasets. The method can be applied to all datasets containing healthy skulls, without the necessity of manual preprocessing and defect synthesis. Therefore, in future work, we plan to acquire and combine more open datasets consisting of healthy skulls from computed tomography.

The disadvantage of the proposed method is a longer training time. Training of the method on just 214 cases required about 1200 epochs to converge, in contrast to the supervised segmentation architectures which all converged before reaching 500 epochs. It is connected with the fact that the proposed method needs to learn the variability of the whole skull, including facial regions and significantly larger defects, not present in the ground truth of the SkullFix/SkullBreak datasets. Nevertheless, there is no significant difference in the inference time between the approaches,  allowing one to perform the inference in less than 100ms using a modern workstation equipped with e.g. NVIDIA RTX 3090 GPU.

The method could be further improved by augmenting the healthy skulls, before performing the patch generation and masking. This could lead to a further increase in the dataset heterogeneity and the model generalizability.

\section{Conclusions}

In this work, we proposed an alternative method to automatic cranial defect reconstruction based on the self-supervised deep masked autoencoders, enhanced by random elastic deformations of the masked input patches. We compared the proposed method to several state-of-the-art solutions and confirmed that it improves the reconstruction quality by more than 0.1 and 1.0 mm in terms of Dice score and Hausdorff distance respectively. In further work, we plan to extend the proposed approach by providing augmentation to the healthy skulls themselves to further increase the dataset heterogeneity. Moreover, we plan to extend the comparison using other datasets and confirm the usability in real clinical settings.


\bibliographystyle{IEEEbib}
\bibliography{refs}

\end{document}